\documentclass[twocolumn,preprintnumbers,amsmath,nofootinbib,amssymb]{revtex4-1}

\usepackage{amsmath,amssymb,epsfig,graphicx}

\usepackage{dcolumn}

\def\be{\begin{eqnarray}}
\def\ee{\end{eqnarray}}
\def\bea{\begin{eqnarray}}
\def\eea{\end{eqnarray}}

\begin{document}

\preprint{}

\title{
Compactification of 6d minimal SCFTs on Riemann surfaces
}

\author{
Shlomo S. Razamat$^1$ and Gabi Zafrir$^2$
}

\

\affiliation{
$^1$Department of Physics, Technion, Haifa 32000, Israel\\ 
$^2$IPMU,  University of Tokyo,  Kashiwa, Chiba 277-8583, Japan\\
}

\date{\today}

\begin{abstract}
We study compactifications on Riemann surfaces with punctures  of  ${\cal N}=(1,0)$ 6d SCFTs with a one dimensional tensor branch and no continuous global symmetries. The effective description of such models on the tensor branch is in terms of pure gauge theories with decoupled tensor. 
 For generic Riemann surfaces, the resulting theories in four dimensions are expected to have ${\cal N}=1$ supersymmetry. We compute the anomalies expected from the resulting 4d theories by integrating the anomaly polynomial of the 6d theory on the Riemann surface. For the cases with $6d$ gauge models with gauge groups $SU(3)$ and $SO(8)$ we further propose a field theory construction for the resulting 4d theories. For the 6d $SU(3)$ theory, we argue that the theories in four dimensions are quivers with $SU(3)$ gauge nodes and free chiral fields. The theories one obtains from the 6d $SO(8)$ gauge theory are quivers with $SU(4)$ gauge groups and chiral fields with R charge a half. In the last case the theories constructed for general Riemann surfaces involve gauging of symmetries appearing at strong coupling.  The conformal manifolds of the models are constructed from gauge couplings and baryonic superpotentials. We support our conjectures by matching the dimensions of the conformal manifolds with complex structure moduli of the Riemann surfaces, matching anomalies between six and four dimensions, and checking the dualities related to different pair of pants decompositions of the surfaces.
 As a simple application of the results we conjecture that $SU(3)$ gauge theory with nine flavors in four dimensions has a duality group acting on the seven dimensional conformal manifold which is the mapping class group of sphere with  ten marked points.

\end{abstract}


\maketitle

\section{Introduction}

\

In recent years a lot of the properties of four dimensional supersymmetric theories were either elucidated or discovered by realising them as compactifications of six dimensional $(1,0)$ models. Following the seminal work of Gaiotto \cite{Gaiotto:2009we} realising ${\cal N}=2$ theories in four spacetime dimensions as compactifications of the 6d $(2,0)$ theories, several compactification scenarios featuring less supersymmetric setups were discussed. These include compactifications of the 6d $(2,0)$ theories with flux for global symmetry  \cite{Benini:2009mz,bahbbwe}, orbifolds of the $(2,0)$ theory  \cite{Gaiotto:2015usa,   Razamat:2016dpl},  the E-string SCFT and more general conformal matter in six diemsnions \cite{Kim:2017toz,Kim:2018bpg,Kim:2018lfo}.

For a generic Riemann surface, Lagrangian descriptions for the resulting theories are for most of the above mentioned cases unknown. The exclusive list of cases where this is known are the $A_1$ and $A_2$ $(2,0)$ theories, the rank $1$ E-string, and the SCFT living on $2$ M$5$-branes in the presence of a ${\mathbb C}^2/{\mathbb Z}_2$ singularity. A common thread in all these theories is that they have only a handful of degrees of freedom in 6d. The same results for their generalizations, by adding tensors, vectors or hypermultiplets, are not known. Furthermore, for the $A_1$ $(2,0)$ theory, the 4d theories are all completely Lagrangian with weak coupling limits, a property that is lost upon adding matter, like the generalization to the $A_2$ $(2,0)$ theory or the SCFT living on $2$ M$5$-branes in the presence of a ${\mathbb C}^2/{\mathbb Z}_2$ singularity.   These theories are constructed  using Lagrangian constructions which involve gauging symmetries appearing in strong coupling cusps of conformal manifolds and thus are inherently strongly coupled.\footnote{One can also construct Lagrangians for theories obtained in compactifications with very special choices of surfaces and punctures  for a wider variety of choices of six dimensional theories \cite{Gaiotto:2009we, Gaiotto:2015usa, Maruyoshi:2016tqk, Kim:2017toz,Kim:2018bpg}.}

This expresses the, at least intuitive, expectation that simpler 6d theories should lead to simpler 4d theories. Of course, this may just be an artifact of our present understanding, soon to change as we acquire more knowledge. Nevertheless, we can try to look at this as a guiding principle. Particularly, this suggests that it may be worthwhile, in the study of compactifications of 6d SCFTs, to look at particularly simple 6d SCFTs with minimal matter content. In the study of the classification of 6d SCFTs, there is a class of minimal models, which are the minimal SCFTs possible with a single tensor multiplet and no global symmetry.

 These SCFTs require vector multiplets in addition to the single tensor, and have a low-energy effective description away from the origin of the tensor branch as a pure 6d gauge theory. The list of 6d anomaly free pure gauge theories is  limited  to the following \cite{Seiberg:1996qx,BeV}: $SU(3)$, $SO(8)$, $F_4$, $E_6$, $E_7$, and $E_8$, and unsurprisingly, these are precisely the theories that appear for the case at hand.\footnote{The $E_7$ gauge theory can also have a single half-hyper in its fundamental representation, and there is a 6d SCFT having this gauge theory as its low-energy effective description on the tensor branch. This then gives another 6d SCFT with a single tensor multiplet and no global symmetry. However, we shall not consider it here.}

The above suggests that it may be interesting to study the compactifications of these theories on Riemann surfaces to 4d, which is what we initiate in this article. Specifically, we concentrate on the cases of 6d gauge group $SU(3)$ and $SO(8)$, and find a conjecture for the theories one obtains in four dimension upon compactification on general Riemann surface with various twisted punctures. From our conjectures we will derive interesting implications about the field theories in four dimensions. In particular these models have no supersymmetric relevant deformations in general, no flavor symmetry on general locus of the conformal manifold,   and will have conformal manifolds matching the complex structure moduli of the compactification Riemann surface. The mapping class group of the complex structure moduli then is expected to be manifest in four dimensions as duality of the models.

While our results are about the relations between the 4d and 6d theories, they also have implications to 5d physics. Particularly, our results suggest the existence of certain 5d description of the above 6d SCFTs compactified on a circle. One of these relations, for the 6d $SU(3)$ gauge theory, was previously proposed in \cite{Jefferson:2018irk}, while the other is to our knowledge, new.

\

\section{Six dimensions}

\

Consider a 6d $\mathcal{N} = (1,0)$ pure gauge theory with a  gauge group $G$. The possible choices for $G$ are severely limited due to the requirement of anomaly cancellation. Specifically, the $\mathcal{N} = (1,0)$ vector multiplet contains a chiral fermion which leads to a gauge anomaly quartic in the field strength $F$. This anomaly comes in two variants, differing by the structure of the index contractions. These are conventionally denoted as $Tr(F^4)$, where the gauge indices are contracted with the quartic Casimir, and $Tr(F^2)^2$, where the gauge indices are contracted with the quadratic Casimir twice. The latter can be solved by the addition of a $\mathcal{N} = (1,0)$ tensor multiplet via the Green-Schwarz mechanism provided it is negative    \cite{Seiberg:1996qx}, which is always the case for the chiral fermion coming from the vector multiplet. Thus, the major constraint is that the $Tr(F^4)$ contribution from the adjoint fermion be zero.

The types of $G$ for which the constraint of anomaly cancellation is obeyed are: $SU(3), SO(8), F_4, E_6, E_7$ and $E_8$. All of these except $SO(8)$ obey the constraint simply as they lack a quartic Casimir and so this type of anomaly does not exist for them.\footnote{The groups $SU(2)$ and $G_2$ also do not have a quartic Casimir. However, these have a Witten type anomaly which excludes the pure case. Incidentally, $SU(3)$ also has that type of anomaly, since $\pi_6 (SU(3)) = {\mathbb Z}_6$, but the contribution of a chiral fermion in the adjoint representation to it turns out to be vanishing \cite{BeV}.} The $SO(8)$ case is in some sense the opposite example, as $SO(8)$ is the only group that has
 more than one quartic Casimir. Particularly, $SO(8)$ has two quartic Casimirs, related to the Euler and second Pontryagin classes, and so has two independent quartic anomalies. The group $SO(8)$ also has an $S_3$ outer automorphism group, generating the so called triality transformations, under which the two quartic Casimirs transform as the irreducible two dimensional representation of $S_3$. Thus,  that must also act linearly on the two $Tr(F^4)$ anomalies. As a result, a sufficient condition for an $SO(8)$ gauge theory to be $Tr(F^4)$ anomaly free, is for it to be triality invariant. Particularly, pure $SO(8)$ gauge theory is triality invariant and so cannot have any $Tr(F^4)$ type anomalies.

For all of these cases, we expect there to be a 6d SCFT with a one dimensional tensor branch (usually referred to as rank $1$ theory), which flows to the pure $G$ gauge theory plus a decoupled tensor on a generic point on the tensor branch. The resulting SCFT should have no continuous flavor symmetries. We can evaluate the 't Hooft anomalies of these SCFTs, using the gauge theory description, and collect the results in an anomaly polynomial $8$-form. This will be useful later when we consider reductions of these SCFTs to 4d.

The 't Hooft anomalies receive contributions from three sources: the chiral fermion in the vector multiplet, the self-dual tensor and chiral fermion in the tensor multiplet, and the Green-Schwarz term required to cancel the $Tr(F^2)^2$, mixed gauge-global and mixed gauge-gravity anomalies. The contributions of the first two are straightforward to write down. Particularly, the vector multiplet of the group $G$ contributes to the anomaly polynomial \cite{Ohmori:2014kda}:

\bea
&-& \frac{Tr(F^4)_{Adj}}{24} - \frac{C_2 (R) Tr(F^2)_{Adj}}{4} -\frac{d_G C^2_2 (R)}{24}\\ &-&\frac{d_G C_2 (R) p_1 (T)}{48}  -  \frac{Tr(F^2)_{Adj} p_1 (T)}{48}\nonumber\\& - & d_G \frac{(7p^2_1 (T) - 4 p_2 (T))}{5760} .\nonumber
\eea   
Here we  use $C_2 (R)$ for the second Chern class of the $SU(2)$ R-symmetry bundle in the doublet representation, and $p_1 (T), p_2 (T)$ for the first and second Pontryagin classes of the tangent bundle respectively. The constant $d_G$ stands for the dimension of the group $G$. In all the cases we consider here $Tr(F^4)_{Adj}$ can be expressed as:

\be
Tr(F^4)_{Adj} = \lambda_G Tr(F^2)^2_{Adj},
\ee 
where we have written the value of the constant $\lambda_G$ in table \ref{Gdata}.

\begin{table}
\begin{center}
\begin{tabular}{|c||c|c|c|c|c|c|}
\hline 
$G$ & $SU(3)$ & $SO(8)$ & $F_4$ & $E_6$ & $E_7$ & $E_8$ \\
\hline
$d_G$ & $8$ & $28$   &  $52$ &  $78$ & $133$ &  $248$   \\
\hline
$\lambda_G$ & $\frac{1}{4}$ & $\frac{1}{12}$   &  $\frac{5}{108}$ &  $\frac{1}{32}$ & $\frac{1}{54}$ &  $\frac{1}{100}$ \\
\hline    
\end{tabular}
 \end{center}
\caption{Group data used in writing down the anomaly polynomial.}
\label{Gdata}
\end{table}

The tensor multiplet contributes:

\be
\frac{C^2_2 (R)}{24} +\frac{C_2 (R) p_1 (T)}{48} + \frac{(23p^2_1 (T) -116 p_2 (T))}{5760} .
\ee

Finally to cancel all gauge anomalies we need to introduce the Green-Schwarz term:

\be
\frac{\lambda_G}{24} (Tr(F^2)_{Adj} + \frac{3}{\lambda_G} C_2 (R) + \frac{1}{4\lambda_G} p_1 (T))^2 .
\ee

Summing up all the terms, we find:

\bea
I^{6d} & = & \frac{1}{24} (\frac{9}{\lambda_G} - d_G + 1) C^2_2 (R) \nonumber \\
&+& \frac{1}{48} (\frac{3}{\lambda_G} - d_G + 1) C_2 (R) p_1 (T) \nonumber \\ & + & \frac{(\frac{15}{\lambda_G} -7 d_G + 23) p^2_1 (T) + (4 d_G -116) p_2 (T)}{5760} \label{AnomPol} . 
\eea     

We next consider compactifying the 6d SCFT to 4d on a Riemann surface. As the Riemann surface is curved, supersymmetry is broken. To avoid this, we twist the $SU(2)$ R-symmetry bundle so as to cancel the curvature of the Riemann surface for some of the supercharges which are charged under it. In such a way we can preserve at most $4$ supercharges, corresponding to $\mathcal{N} = 1$ in 4d. The twist also breaks the $SU(2)$ R-symmetry to its $U(1)$ Cartan, which becomes an R-symmetry in 4d. We can predict the 4d anomalies of the resulting theory from the anomaly polynomial of the 6d SCFT, by integrating the latter on the Riemann surface \cite{Benini:2009mz}. This should lead to the anomaly polynomial of the 4d theory, which contains the 't Hooft anomalies of the 4d theory, at least those for symmetries descending from 6d.  

In our case, we need to integrate (\ref{AnomPol}) on the Riemann surface, but first we need to take the twist into account. This is done by setting $C_2 (R) = -C_1 (R)^2 + 2 (1-g) t C_1 (R) + ...$, where $C_1 (R)$ is the first Chern class of the $U(1)$ Cartan of the $SU(2)$ R-symmetry and $t$ is a unit flux 2-form on the Riemann surface. Inserting this into (\ref{AnomPol}) and integrating we find:
\bea
I^{4d} & = & \frac{1}{6} (\frac{9}{\lambda_G} - d_G + 1) (g-1) C^3_1 (R) \\ \nonumber & - & \frac{1}{24} (\frac{3}{\lambda_G} - d_G + 1) (g-1) C_1 (R) p_1 (T). 
\eea  

From this we can evaluate the conformal central charges finding:

\bea
&& a = \frac{3}{16} (\frac{12}{\lambda_G} - d_G + 1) (g - 1), \\ \nonumber
 \;\;\;\, && c = \frac{1}{8}(\frac{33}{2\lambda_G} - d_G + 1) (g - 1) .
\eea

We can also consider compactifications on surfaces with punctures. To do so we need to supply some information at the punctures. A convenient way to achieve this is to elongate the region near a puncture to a cylinder. The effective theory on this cylinder is given by the five dimensional reduction of the theory. Therefore, to make progress with considering punctures we need to consider the 6d SCFTs on a circle.

\

\noindent{\bf {Reduction to 5d}}

\

Here we consider the previously discussed 6d SCFTs reduced on a circle to 5d.\footnote{For a previous study of this problem see \cite{DelZotto:2015rca}.} It is by now well appreciated that reducing 6d SCFTs on a circle, possibly with holonomies for flavor symmetries, can lead to effective 5d gauge theory descriptions, where the 5d gauge couplings being related to the holonomies and the compactification radius. Here, as the 6d SCFTs have no continuous global symmetry, we cannot incorporate holonomies for flavor symmetries. 

As a result, when we compactify the 6d SCFT on a circle, if we are to get a gauge theory it can only be a pure 5d gauge theory which has the 6d theory as its UV completion. The basic reason is that in 5d we can only have a single $U(1)$ global symmetry, associated with the compactification, and as any 5d gauge theory comes with a topological $U(1)$, we are limited to pure 5d gauge theories. Most pure 5d gauge theories are known to have a UV completion as 5d SCFTs and so cannot be used for this purpose. Recently a series of criteria for a 5d gauge theory to have a 5d or 6d SCFT UV completion was proposed in \cite{Jefferson:2017ahm}, and its implications studied for the case of single gauge groups. While the proposed criteria are known not to be sufficient \cite{Jefferson:2018irk}, it is believed to be necessary, and for the purpose of this article, we shall assume that it is indeed so.

Going over the theories listed in \cite{Jefferson:2017ahm}, we find that there are only three pure 5d gauge theories that can have a 6d SCFT as  their minimal UV completion. These are $SU(3)$ with Chern-Simons $\pm 9$, $SU(4)$ with Chern-Simons $\pm 8$, and $SU(6)$ with Chern-Simons $\pm 9$. Therefore, the majority of the 6d SCFTs of the class we discussed do not posses 5d gauge theory descriptions. This means that it is unclear how punctures can be defined in that case, and we will not consider them here. Instead we shall concentrate on the three cases that may have a 5d gauge theory description.

Out of these three, only the $SU(3)$ gauge theory with Chern-Simons level $\pm 9$ is known to exist as it has been constructed through geometrical means in \cite{Jefferson:2018irk}. The authors of \cite{Jefferson:2018irk} also provide evidence that this 5d gauge theory lifts to the 6d SCFT completion of the pure 6d $SU(3)$ gauge theory with a twist. More specifically, the 6d $SU(3)$ gauge theory has  a discrete symmetry given by charge conjugation, and assuming it lifts to a symmetry of the SCFT, we can consider a compactification with a twist under this discrete symmetry. The proposal in \cite{Jefferson:2018irk} then, is that when the 6d SCFT completion of the pure 6d $SU(3)$ gauge theory is compactified on a circle with this twist, the resulting low-energy theory is the 5d $SU(3)$ gauge theory with Chern-Simons level $\pm 9$. Note that the twist is necessary, for instance, to make the Coulomb branch dimensions agree. On the 5d side we have the two dimensional Coulomb branch of $SU(3)$ which should match the two dimensional Coulomb branch of the 6d $SU(3)$ on a circle plus the one dimensional tensor. However, the twist projects the dimension three Coulomb branch operator of the 6d $SU(3)$ on a circle, as it is odd under it, making the matching work.

We shall later on see that 
our 4d results are consistent with this picture. Particularly, we will construct theories whose anomalies match those of the 6d $SU(3)$ SCFT on Riemann surfaces with punctures possessing $SU(3)$ global symmetry and with anomalies matching those evaluated by assuming the 5d description is an $SU(3)$ gauge theory with Chern-Simons level $\pm 9$. Similarly, we shall construct theories whose anomalies match those of the 6d $SO(8)$ SCFT on Riemann surfaces with punctures possessing $SU(4)$ global symmetry and with anomalies matching those evaluated by assuming the 5d description is an $SU(4)$ gauge theory with Chern-Simons level $\pm 8$. While this is indirect evidence, it does suggests that, first, the $SU(4)$ gauge theory with Chern-Simons level $\pm 8$ exists as a 6d lifting theory, and, second, that it lifts to the 6d $SO(8)$ SCFT. Note, however, that the Coulomb branch dimension of the 5d gauge theory does not match that of the 6d SCFT on a circle unless some directions are projected by a twist, like in the $SU(3)$ case. Fortunately, the 6d $SO(8)$ gauge theory has a discrete $S_3$ symmetry given by the outer automorphisms of $SO(8)$, and assuming these lift to symmetries of the SCFT, they can be used to form twisted compactifications. The ${\mathbb Z}_3$ element, implementing the $SO(8)$ triality, in particular, when used in twisted compactification, projects the two dimension four Coulomb branch operators of the 6d $SO(8)$ gauge theory on a circle. Thus, from these considerations, we see that if the picture we suggest is to be correct the compactification must be with at least this ${\mathbb Z}_3$ twist. Finally, we do not have theories that can be associated with the 5d $SU(6)$ gauge theory with Chern-Simons level $\pm 9$, and so have nothing to say about this case.

We will next explore some of the implications  of the 5d descriptions for the cases we consider. It should be noted that ultimately we will be concerned with the 4d theories resulting from the 6d SCFTs. There are various aspects, particularly related to twists, that are difficult to decipher just from the 4d theories. As a result, there are various puzzles that we cannot resolve at this moment. We shall first ignore the twist, and concentrate on those elements that lead to sharp 4d predictions that can be compared against the 4d theories. These particularly are the contribution of the punctures to the 't Hooft anomalies.    

We shall define the punctures by boundary conditions for the 5d fields. We are interested in preserving four supercharges and a natural boundary condition is to choose Dirichlet for the vector fields and Neumann for the adjoint chiral, where here we are using 4d $\mathcal{ N  } = 1$ language. This will imply that we have Dirichlet boundary condition for chiral fermions and Neumann for antichiral. From this we can compute the anomaly inflow contribution of the puncture to four dimensions. This comes from the anti-chiral fermions and the CS term. The former are in the adjoint of the 5d gauge groups, $SU(3)$ or $SU(4)$ depending on the case, have R charge $-1$, and contribute half of the contribution of four dimensional fermions.  The gauge symmetry in five dimensions will be frozen and becomes a flavor symmetry.
 For the 5d $SU(3)$ gauge theory, which we associate with the 6d $SU(3)$ SCFT, This gives the anomaly contributions $Tr (R)=-4$, $Tr\, (R^3)=-4$, and $Tr (R SU(3)^2)=-\frac32$ coming from every puncture. To compute the full anomaly polynomial we also need to take into account the computation of the integral of the 6d anomaly polynomial on the Riemann surface which is the same as above but with $g-1\to g-1+s/2$ where $s$ is the number of punctures, and we specialize to the case of the 6d $SU(3)$ SCFT. These give for surfaces with punctures the anomalies:
\be
&&a=\frac3{32}(82(g-1)+33s),\;\;\;\;\,\\&& c=\frac1{16}(118(g-1)+51s)\, .\nonumber
\ee  In addition the CS term contributes $Tr (SU(3)^3) =\pm 9$, where the sign is determined by the sign of the CS term. 
 
A similar story also holds for the 5d $SU(4)$ gauge theory, which we associate with the 6d $SO(8)$ SCFT. The puncture is defined using the same boundary  conditions for the 5d fields as previously considered.  This gives the anomaly contributions $Tr (R)=-\frac{15}2$, $Tr\, (R^3)=-\frac{15}2$, and $Tr (R SU(4)^2)=-2$ coming from every puncture. To compute the full anomaly polynomial we also need to take into account the computation of the integral of the 6d anomaly polynomial on the Riemann surface and the inflow. These give for surfaces with punctures the anomalies:
\be
&&a=\frac9{16}(39(g-1)+17s),\;\;\;\;\,\\&& c=\frac3{8}(57(g-1)+26s)\, .\nonumber
\ee  In addition the CS term contributes $Tr (SU(4)^3) =\pm8$, where the sign is determined by the sign of the CS term. 

\

\noindent{\bf {Implications of the twists}}

\

Before moving on we are going to consider some of the implications of the twists, that we have argued should be there from reasoning associated with the 5d reduction. The incorporation of twists is something that is familiar from class ${\cal S}$ \cite{Gaiotto:2009hg,Gaiotto:2009we} constructions and many of the expectations we mention here have analogues there as well  \cite{CDTAt,CDTDN,CDTn}. In those cases the twist can be turned on along any cycle on the surfaces. These particularly include those surrounding the punctures. For the special case of a sphere with $q$ punctures, these are the only cycles. Furthermore, they are not independent, where the  holonomy around $q-1$ punctures should be the inverse of that around the remaining puncture. In our case, since the punctures we consider must incorporate a twist, this leads to constraints on the possible theories. For instance if the punctures have a ${\mathbb Z}_2$ twist then we cannot have a sphere with odd number of punctures. Also one can only glue a twisted puncture to another twisted puncture with opposite twist.

On Riemann surfaces without punctures, the twists can still be incorporated on the cycles of the Riemann surface. When the latter is built from punctured spheres then whenever we glue two twisted punctured we get the associated twist along the cycle spanned by the punctures. Additionally, we also have the other tube cycle, which is incorporated by how the gluing is performed. Thus, in theories supporting the twist there are several ways to perform the gluing, differing by whether one incorporate a twist or not. For instance in the case of the A type $(2,0)$ theory, this distinction is given in the gluing by whether the fundamental representation of the $SU$ group associated with one puncture is mapped to the fundamental or anti-fundamental representation of the other puncture. It should be noted that such twists around cycles do not affect the anomalies save by possibly breaking symmetries, which does not occur in the cases we consider here as these have no global symmetries. As a result it may be difficult to distinguish between 4d theories resulting from compactifications differing by application of twists.  

We next wish to consider each case in turn. We begin with the 6d $SU(3)$ SCFT. As we previously mentioned, when compactified to 5d on a circle with a twist, it is expected to give the $SU(3)$ gauge theory with Chern-Simons level $\pm 9$. One issue with this picture is that the 5d gauge theory comes in two inequivalent variants differing by the sign of the Chern-Simons level, but it is not clear what degree of freedom in 6d determines to which of this version one would flow. One possibility is that the discrete symmetry group of the SCFT is larger than ${\mathbb Z}_2$ and that there are then different twists accounting for this difference.\footnote{We thank Yuji Tachikawa for a discussion regarding this point.} We do observe in 4d that we only find theories related to spheres with an even number of punctures, which is consistent with a ${\mathbb Z}_2$, but also with possibly other twist if the punctures come in pairs with opposite twists. As a result we cannot resolve this ambiguity and we leave this for future study.

We next move to consider the 6d $SO(8)$ SCFT. As we previously mentioned, when compactified to 5d on a circle with a twist, it is expected to give the $SU(4)$ gauge theory with Chern-Simons level $\pm 8$. Here, since the twist is ${\mathbb Z}_3$ we have two conjugate twists and the previously mentioned issue does not arise. Also here in 4d we do find a theory that can be associated with a three punctured sphere, which is consistent with punctures having a ${\mathbb Z}_3$ twist.   

\

\section{Four dimensions}

\

\noindent{$\mathbf{SU(3)}$}

\

We want to find four dimensional models such that they will have factors of $SU(3)$ global symmetry and the anomalies of these symmetries are $Tr(SU(3)^3)=\pm 9$, $Tr (R SU(3)^2)=-\frac32$. Let us make a guess that we can do this with free fields. This guess does not have to be correct but we will see it works in the case we study.
We  discover that constructing field theories such that we will have nine free chiral fields in the fundamental of $SU(3)$ produces the right anomalies. The question is then how we build models from these collections of fields. It is not hard to come up with the following conjectures which reproduce all the robust expectations we have derived from pure $SU(3)$ gauge theory in six dimensions.

\begin{figure}[tbph]
\includegraphics[scale=0.82]{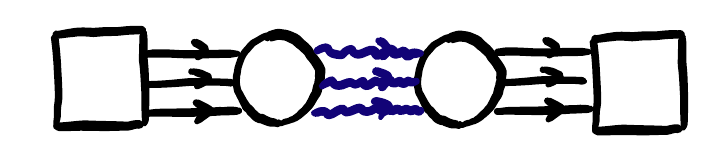}
\caption{The sphere with four punctures. The squares are $SU(3)$ flavor and circles $SU(3)$ gauge. We have additional $SU(3)$ global symmetries rotating the three bifundamentals. The bifundamentals in the middle link have a baryonic superpotential turned on breaking the $SU(3)$ flavor symmetry. The wavy lines denote the fact that the fields have the baryonic superpotential. All in all we are left with four $SU(3)$ symmetries. }
\end{figure}

\

\noindent{\bf {The conjectures}}
We conjecture that the theory corresponding to a sphere with four punctures is constructed from three sets of trifundamental $SU(3)$ chiral fields, gauging two diagonal $SU(3)$s  of two different  trifundamentals, and breaking the additional $SU(3)$ by a superpotential. The model then has four $SU(3)$ factors of global symmetry.  See Fig. 1 for details.

Let us denote the three trifundamental fields as $Q^{(i)}_{abc}$. The superpotential is,
\be
\sum_{d=1}^3\lambda_d \, \epsilon^{abc}\epsilon^{efg} Q^{(2)}_{ade}Q^{(2)}_{bdf}Q^{(2)}_{cdg}\,.
\ee 
This superpotential is marginally irrelevant near the free locus and we will discuss the meaning of this later.
We can combine the four puncture spheres to form arbitrary surfaces by gauging diagonal combinations of $SU(3)$ global symmetries with ${\cal N}=1$ vector multiplets. As each such group has nine flavors the one loop beta function is zero. It is convenient to define building blocks of Fig. 2 to write down the quivers for such theories. The four puncture sphere is written using the blocks in Fig. 3 and  Fig. 4 shows an example of a more general surface.
\begin{figure}[htbp]
\includegraphics[scale=0.62]{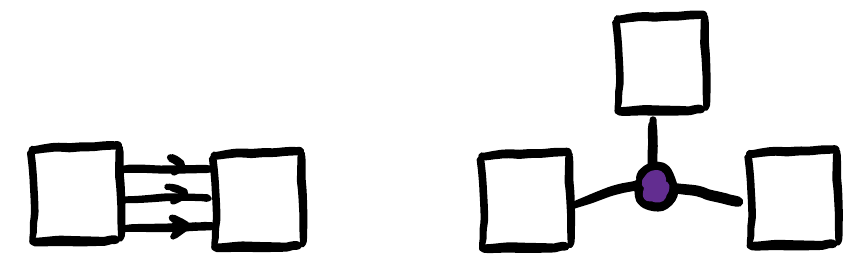}\; \;\, \includegraphics[scale=0.62]{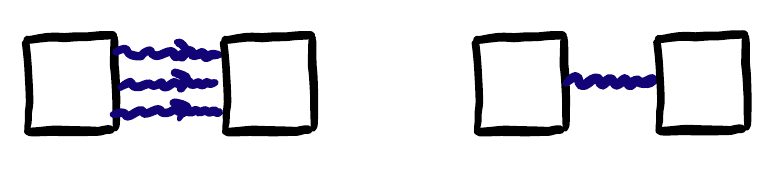}
\caption{Building blocks of the theories. First we have the trifundamental, and then trifundamental with one of the three $SU(3)$ symmetries broken by baryonic superpotential which preserves other two $SU(3)$ flavor symmetries.}
\end{figure}
It is easy to verify that with these blocks the 't Hooft anomalies match exactly the anomalies computed from six dimensions. The anomaly of the four punctured sphere is $a=45/16$ and $c=43/8$. We glue surfaces with $SU(3)$ ${\cal N}=1$ gauging. We can construct genus $g$ surface with $s$ punctures by gluing together $g+s/2-1$  four punctured spheres with $2g-2+s/2$ gluings. Note that number of punctures is even. 
\begin{figure}[htbp]
\includegraphics[scale=0.62]{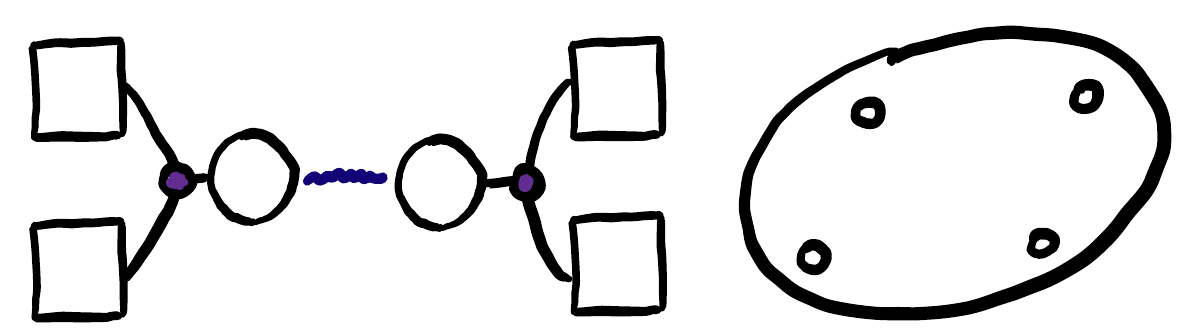}
\caption{Four puncture sphere}
\end{figure} 

We also note that the gluing seems to be determined from anomaly cancellation, and so we do not see any additional option that would incorporate the twist on the cycle of the tube used in the gluing. It is possible that the other choices are more complicated, or maybe this 6d option has no effect on the 4d physics. This also means that we cannot fully pinpoint what are the twists along these tube cycles for the 4d theories we present, as anomalies cannot distinguish this. For the cycles running along the punctures, the twist can be determined from those associated with the punctures. 

Another interesting check is to compute the supersymmetric index which will tell us what are the relevant and marginal operators. 
The supersymmetric index \cite{Romelsberger:2005eg} is defined as the trace over the states of the model in radial quantization,
\be 
Tr (-1)^F q^{j_1-j_2+\frac12 R} p^{j_1+j_2+\frac12 R} \prod_{j=1}^{Rank\, ( G )} a_j^{q_f}\,, 
\ee 
where for states contributing to the index, the conformal dimension is determined in terms of the rest of the charges, $E-2j_1-\frac32 R=0$.
Here $F$ is the fermion number, $j_i$ are the Cartan generators of the $su(2)\times su(2)$ Lorentz symmetry, $E$ is the scaling dumension, $R$ is the R symmetry, $G$ is the flavor group with $q_i$ being charges under the Cartan of it. We use the superconformal R symmetry to compute the index and then one can argue  that   the scalar components of chiral operators appear with powers $(qp)^{\frac{R}2}$. In expansion of the index of an interacting theory in powers of $q$ and $p$ the operators contributing with powers of $qp$ less than  two  come from relevant chiral operators. At order $qp$ we get marginal operators but also fermions from the conserved currents which contribute with negative sign and are in the adjoint representation of the global symmetry \cite{Beem:2012yn}.
For low genus and small number of punctures the index has special features, some  of which we will discuss,  but for generic surfaces the index is,
\be
1+qp((3g-3+s) +\sum_{j=1}^s (-{\bf 8}_j+{\bf 10}_j))   +   \;\;\, \cdots  \,.
\ee As there are no terms below $qp$, there are no relevant operators. The number of exactly marginal directions preserving puncture symmetries is $3g-3+s$ which is expected to be related to complex structure moduli. We also have a conserved current for each puncture and a set of marginal operators charged under the puncture symmetries.

   \begin{figure}[htbp]
\includegraphics[scale=0.42]{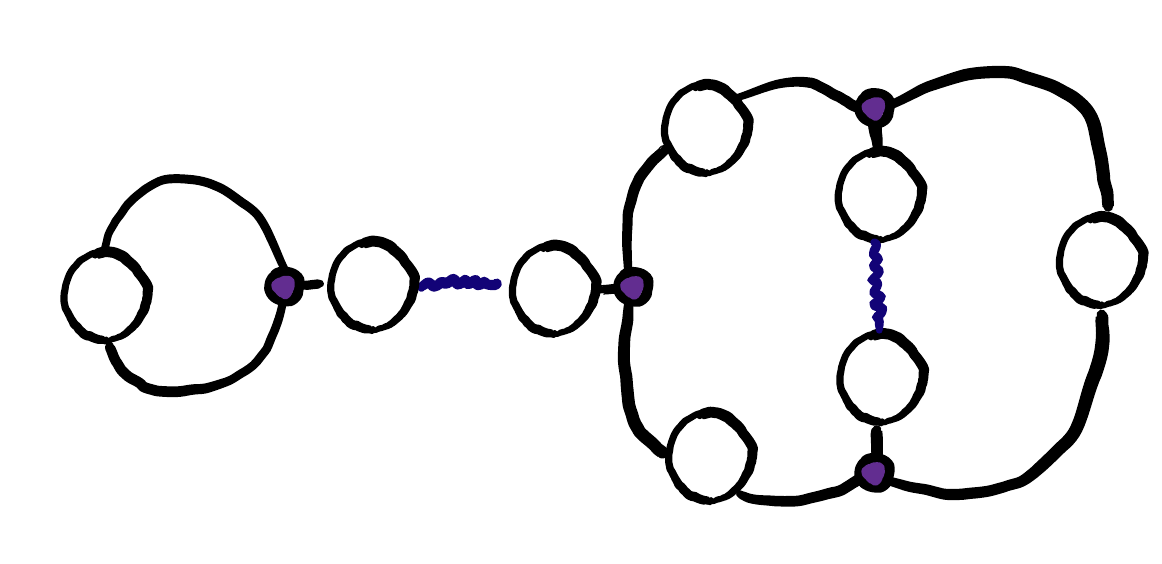}\; \;\, \includegraphics[scale=0.42]{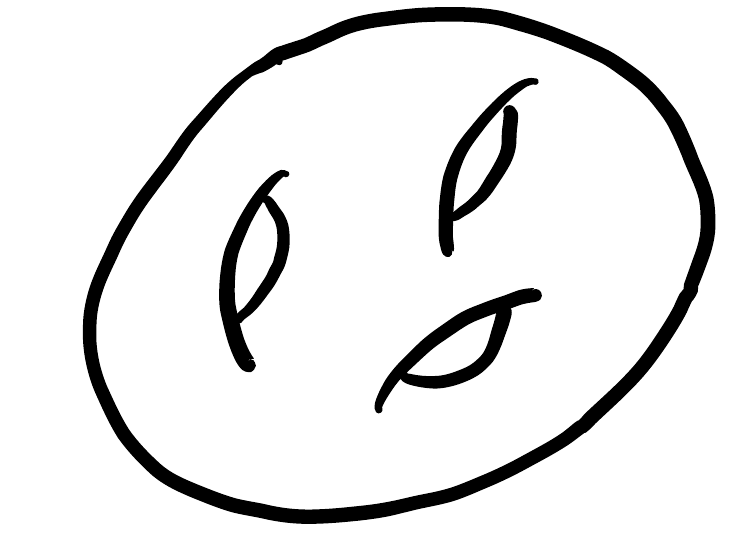}
\caption{Example of genus three model.
 }
\end{figure}

\

\noindent{\bf {Duality}}
For the conjecture to be correct  combining the four puncture spheres to form surfaces of same topology in different ways should give equivalent theories up to the action of dualities. In particular 
the protected spectrum of the theory has to be invariant under exchange of the different factors
 of $SU(3)$ symmetry associated to punctures. This is a non obvious fact. In particular if this holds for the four puncture sphere it will hold for any surface. We have checked this by explicit evaluation of the index in a series expansion. In fact we find that a stronger statement from that which we need appears hold true. Namely, gluing two trifundamentals and ignoring the baryonic symmetry, which is broken for general surfaces on the conformal manifold, the index is invariant under exchanging the four $SU(3)$ symmetries. The index is given by,
 \be
&&\;\;\, G(a^{(b)})= (q;q)^2(p;p)^2\frac16\oint\frac{dh_3}{2\pi i h_3}\oint \frac{dh_2}{2\pi i h_2}\\
&& \frac{\prod_{i,j,k=1}^3
\Gamma_e((q\, p)^{\frac13} a^{(1)}_i a^{(2)}_j h^{-1}_k)\Gamma_e((q\, p)^{\frac13} a^{(3)}_i a^{(4)}_j h_k)}
{
\prod_{i\neq j} 
\Gamma_e(h_i/h_j)}\, ,
\nonumber
\ee  where $\prod_{i=1}^3h_i=\prod_{j=1}^3 a^{(d)}_j =1$ ith $a^{(b)}$ parametrizing the four $SU(3)$ flavors. The above is 
invariant under permutations of the four $a^{(d)}$.  We have checked this in perturbative expansion in the fugacities to order $(qp)^2$. 

Let us concentrate on studying the duality of the four punctured sphere. We can count the dimension of the conformal manifold of the theory near  weak coupling. The full symmetry of the theory is broken on the conformal manifold and the dimensions is ${\it 252}$. Moreover as the superpotential we turn is built from fields having same baryonic charge it is marginally irrelevant at weak coupling. Thus, there is no conformal manifold passing through zero coupling and having the symmetries we are interested to have, four $SU(3)$ and no baryonic symmetry present. However it can happen that at finite position on the conformal manifold we have such a locus of exactly marginal deformations. We can check this by assuming the symmetries we want and computing the index. The index then will tell us what are the marginal operators under these assumption and we can perform the analysis of exactly marginals. It can be that we will get the same answer as at weak coupling, and then the conjecture is consistent, or different, which will prove it not right.
The index assuming the symmetry we are interested in is,
\be
1+q \; p( 3({\bf 3}_1,{\bf 3}_2,{\bf 3}_3,{\bf 3}_4)+1+\sum_{j=1}^4 (-{\bf 8}_j+{\bf 10}_j))+\cdots  .\,\nonumber\\
\ee 
We can break all the symmetry on the conformal manifold as ${\bf 10}$ of $SU(3)$ has non trivial invariants. This means that again we get ${\it 252}$ exactly marginal deformations. This shows that it can be that there is a locus, in fact a line as can be seen from the index, on the conformal manifold such that we have four factors of $SU(3)$ and no baryonic symmetry.

The fact that the relevant locus does not pass through weak coupling is consistent with the fact that we do not know what the sphere with three punctures is. One puncture or all of such a theory should be not twisted, but we do not have a gauge theory description of compactification with no twist. We thus expect the decoupling limit of four punctured sphere to be given by some strongly coupled model.

We expect to have a duality acting on the line of the conformal manifold we find. This duality should permute different symmetry factors.

\

\noindent{\bf {Closing punctures}}
One additional operation we can perform is to study flows which are triggered by vacuum expectation values to operators charged under puncture symmetry. Such flows will break the symmetry of the puncture and will leave us with punctures of different type. As the punctures have twist associated to them  it is impossible then to close the puncture completely.   
Each puncture comes with a marginal operator in the ${\bf 10}$ of $SU(3)$. We can close a puncture by giving a vacuum expectation value to a component of this operator.
A natural choice is to do as follows. First we have that the fundamental of $SU(3)$ has the following character ${\bf 3}=b_1+b_2+b_3$, with $b_1 b_2 b_3 = 1$, and 
\be
{\bf 10} = \sum_{i\neq j} b_i/b_j +\sum_{i=1}^3 b_i^3+1\,.
\ee
 In index notations we give expectation value to operator contributing with weight  $q p b_1^3$ setting this combination of fugacities to one, that is we give a vacuum expectation value to a single bifundamental field.   The effect of the flow is simple, the trifundamental  associated to our puncture is removed from the theory with the symmetry, to which we gave a vacuum expectation value,
               removed and the two additional $SU(3)$ groups identified.  We interpret this procedure as closing a puncture to a different puncture which has no flavor symmetry and we will refer to that as an empty puncture. 
  \begin{figure}[htbp]
\includegraphics[scale=0.62]{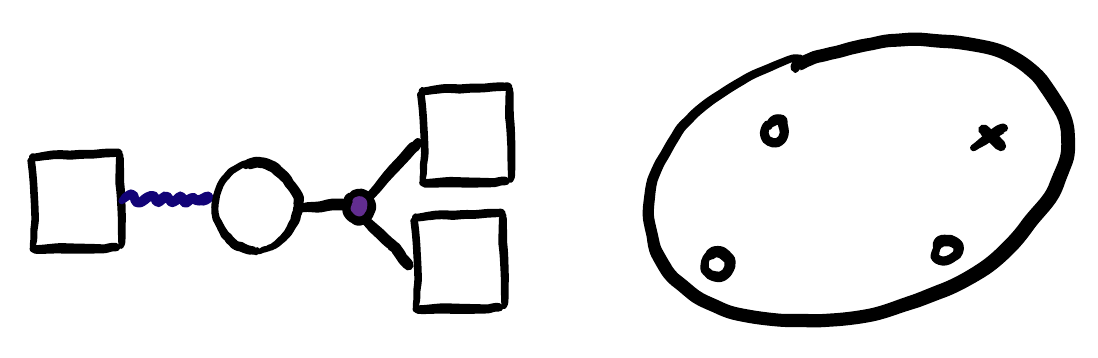}
\caption{Closing a puncture of the four punctured sphere. The circles are $SU(3)$ punctures and crosses empty ones.}
\end{figure}
 Let us add $s'$ empty punctures by starting from a theory with $s'+s$ punctures and closing $s'$ of those. The anomaly is easy to compute and we obtain that empty punctures behave as one third of the $SU(3)$ puncture. We have,
\be
&&a=\frac3{32}(82(g-1)+33s+11s'),\;\;\;\;\,\\&& c=\frac1{16}(118(g-1)+51s+17s')\, .\nonumber
\ee
 We can also compute the index and find that it is,
\be
1+q \; p( 3g-3+s+s'+\sum_{j=1}^s (-{\bf 8}_j+{\bf 10}_j))+\cdots  \;.\,
\ee  This is  consistent with having a conformal manifold preserving the puncture symmetries whose dimension is the same as the space of complex structure moduli. We notice that a theory with $s$ $SU(3)$ and $s'=3n+k$ empty punctures is on the same manifold as the theory with $s+n$ $SU(3)$ punctures and $k$ empty punctures. In particular the same theory is on the same manifold as $s'+3s$ empty punctures. 
  \begin{figure}[htbp]
\includegraphics[scale=0.62]{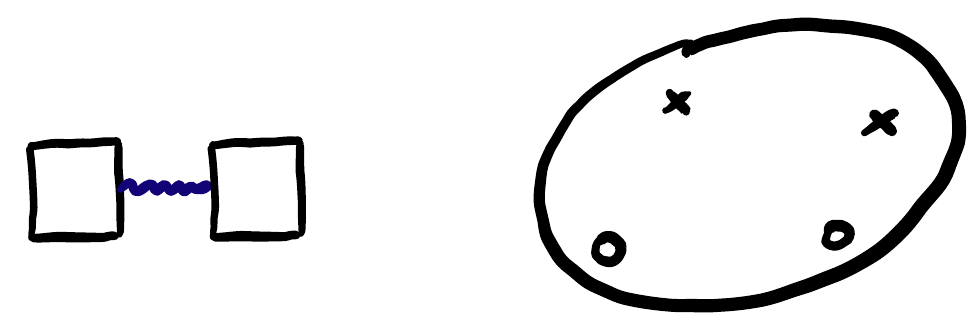}
\caption{Closing $SU(3)$ puncture of Fig. 5 we obtain a theory with two $SU(3)$ and two empty punctures. }
\end{figure}  Note that the theory of Fig. 5 then can be thought off as a sphere with ten empty punctures and has a seven dimensional conformal manifold \cite{Green:2010da,Leigh:1995ep}. We then expect to have the mapping class group of the sphere with ten punctures acting on this conformal manifold.
The cusp with zero coupling corresponds to pair of pants decomposition to two spheres with one $SU(3)$ and five empty punctures.

\

\

\noindent{$\mathbf{SO(8)}$}

\

We want to find four dimensional models such that they will have factors of $SU(4)$ global symmetry and the anomalies of these symmetries are $Tr(SU(4)^3)=\pm 8$, $Tr (R SU(4)^2)=-2$. Let us make again a  guess that we can do this with chiral fields.  The anomalies can be achieved if every $SU(4)$ flavor symmetry has  a single octet of
fundamental chiral fields with $R$ charge half.  We construct the following models reproducing the anomalies and all the predictions of six dimensions.

\

\noindent{\bf {The conjectures }}
For this case the twist is ${\mathbb Z}_3$ and thus we can construct a sphere with three punctures such that all are twisted.
We conjecture that the theory corresponding to a sphere with three punctures is constructed as follows. We define a basic building block to be a pair of bifundamental chiral fields of $SU(4)$. We glue three such blocks linearly together by gauging diagonal $SU(4)$ symmetries. We also turn on baryonic superpotential for the second  block. The symmetry of the model that we identify is two copies of $SU(4)$ at the ends and $SU(2)^2$ rotating the pairs of bifundamentals of $SU(4)$. We conjecture that  the conformal manifold of this model has a special point  so that the $SU(2)^2$ enhances to $SU(4)$. The embedding of $SU(2)^2$ in $SU(4)$ is such that the fundamental ${\bf 4}$ decomposes as $({\bf 2},{\bf 2})$. We then have three $SU(4)$ symmetries. 
One way to check this is to compute the index and see that it is consistent. In particular computing the index of the three puncture sphere and decomposing the $SU(4)$ punctures as $SU(2)\times SU(2)$ we can check that indeed the $SU(2)\times SU(2)$  
  symmetry we see in the action  will appear in the same way as the $SU(2)\times SU(2)$  imbedded in the $SU(4)$ of the puncture.
     We performed this check
      in expansion of the index to low orders in $qp$. It would be interesting to verify whether this holds to all orders. This enhancement of symmetry is a crucial conjecture for our statements  to be correct. In a similar manner to the previous case the baryonic superpotential is marginally irrelevant in the vicinity of the locus preserving the baryonic symmetry. However it is possible that at the locus of the conformal manifold where the symmetry conjecturally enhances to $SU(4)$ this superpotential is turned on.

\

\begin{figure}[htbp]
\includegraphics[scale=0.62]{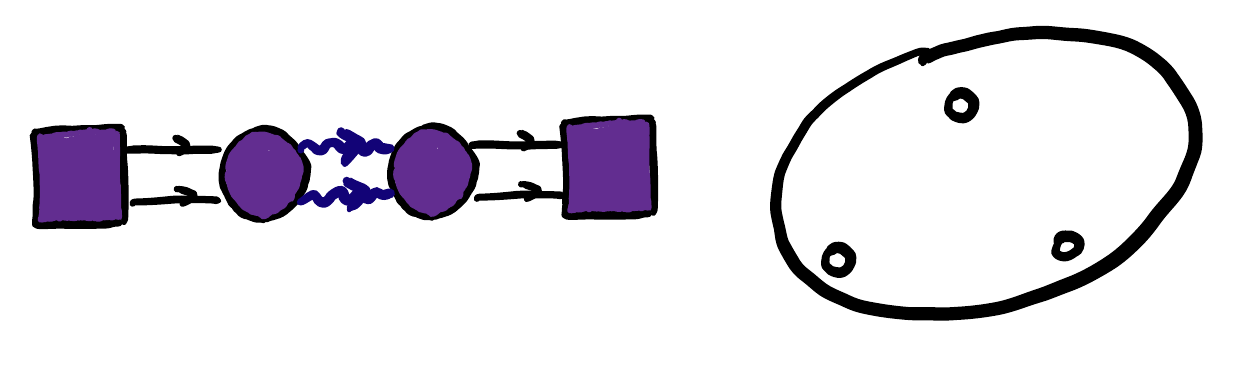}
\caption{The sphere with three punctures. The colored squares are $SU(4)$ flavor and circles $SU(4)$ gauge. We have  $SU(2)$ global symmetries rotating the two bifundamentals. The bifundamentals in the middle link have a baryonic superpotential turned on breaking the $SU(2)$ flavor symmetry. All in all we are left with two $SU(4)$ and $SU(2)^2$ symmetries. The latter is conjectured to be enhanced to $SU(4)$ somewhere on the manifold.}
\end{figure}

To construct the general models corresponding to higher genus surfaces we need to couple all the three $SU(4)$ factors of the three puncture spheres to dynamical vector fields. This involves tuning the coupling to the locus where the symmetry enhances. In this respect the theory is not Lagrangian in the usual sense and similar constructions exist for other setups arising in various compactifications of six dimensional theories to four dimensions \cite{Gadde:2015xta,Razamat:2016dpl,Kim:2017toz,Agarwal:2018ejn}.  We can  write the index translating the $SU(2)\times SU(2)$ representations to $SU(4)$, which can be easily obtained at least in low orders in expansion of the index. The result then is,
\be
1+(q \, p)^{\frac34}2({\bf 4},\, {\bf 4},\, {\bf 4})+ q p( \sum_{i=1}^3(-{\bf 15}_i+{\bf 20}'_i)\,)+\cdots\;\, .\nonumber\\
\ee
Here ${\bf 4}$, ${\bf 15}$, and ${\bf 20}'$ stand for characters of representations of $SU(4)$ details of which will be given later. 
Note that the conformal manifold preserving all the symmetry we want is a point consistent with the theory corresponding to a three puncture sphere.
Using this we  compute the supersymmetric index   for general surfaces, which will tell us what are the relevant and marginal operators,
\be
1+qp((3g-3+s) +\sum_{j=1}^s (-{\bf 15}_j+{\bf 20}'_j))   +   \;\;\, \cdots  \,.
\ee We have $s$ punctures and genus $g$ in this expression. As there are no terms below $qp$, there are no relevant operators \cite{Beem:2012yn}. The number of exactly marginal directions preserving puncture symmetries is $3g-3+s$ which is expected to be related to the complex structure moduli. We also have a conserved current for each puncture and a set of marginal operators charged as ${\bf 20}'$ under the puncture symmetries.

For the conjecture to be true we need to check that the duality exchanging the four factors of the $SU(4)$ symmetry in the four puncture sphere is the symmetry of the index. This is indeed the case at least to low orders in expansion of the index as can be seen above. 

Finally we note that here also the gluing seems to be determined from anomaly cancellation, and again we do not see any additional option that would incorporate the twist on the the cycle of the tube used in the gluing. Likewise, we also have an ambiguity in matching the 4d theories to the 6d compactification regarding
 the exact choice of twist lines.

\

\noindent{\bf {Closing punctures}}
Here we can close the $SU(4)$ puncture down to an $SU(2)$ puncture and then further to a puncture with no symmetry. First we have that ${\bf 4}=b_1+b_2+b_3+b_4$, with $b_1 b_2 b_3 b_4 = 1$, and 
\be
{\bf 20}' = \sum_{i\neq j} b_i/b_j +\sum_{i,j=1,\, i>j}^3 ( b_i^2 b_j^2+\frac1{b_i^2 b_j^2} ) +1+1\,.
\ee
Note also that under the $SU(2)\times SU(2)$ subgroup of $SU(4)$, the ${\bf 4}$ has the following character, $(u+\frac1{u})(\widetilde u+\frac1{\widetilde u})$ and we have
\be
&&{\bf 20}'\to 1+{\bf 5}+\widetilde {\bf 5} +{\bf 3}\,\widetilde {\bf 3}\, ,\,\;\\
&&{\bf 15}\to {\bf 3}+\widetilde {\bf 3} +{\bf 3}\,\widetilde  {\bf 3}\, .\nonumber
\ee
We can turn on a vacuum expectation value for an operator in the ${\bf 5}$ of $SU(2)$ which will trigger a flow ending in a fixed point with the $SU(4)$ puncture traded with a puncture with $SU(2)$ symmetry.  For the quiver theories this is easy to understand. We give an expectation value to a baryonic operator of one of the basic blocks  built from  one of the two bifundamental chiral fields. This will remove the block via the Higgs mechanism. We can do this procedure for the remaining  $SU(2)$ group and get an empty puncture. Because of the twist we cannot remove a puncture completely and are left with a puncture with no symmetry associated 
with it.

The anomalies of a general surface with $s$ $SU(4)$ punctures, $s'$  $SU(2)$ punctures and $s''$ empty punctures are obtained by starting with $s+s''+s'$ $SU(4)$ punctures and giving the needed vaxuum expectation values,
\be
&&a=\frac9{16}(39(g-1)+17s+\frac{34}3s'+\frac{17}3 s''),\;\;\;\;\,\\&& c=\frac3{8}(57(g-1)+26s+\frac{52}3 s'+\frac{26}3 s'')\, .\nonumber
\ee
\begin{figure}[htbp]
\includegraphics[scale=0.48]{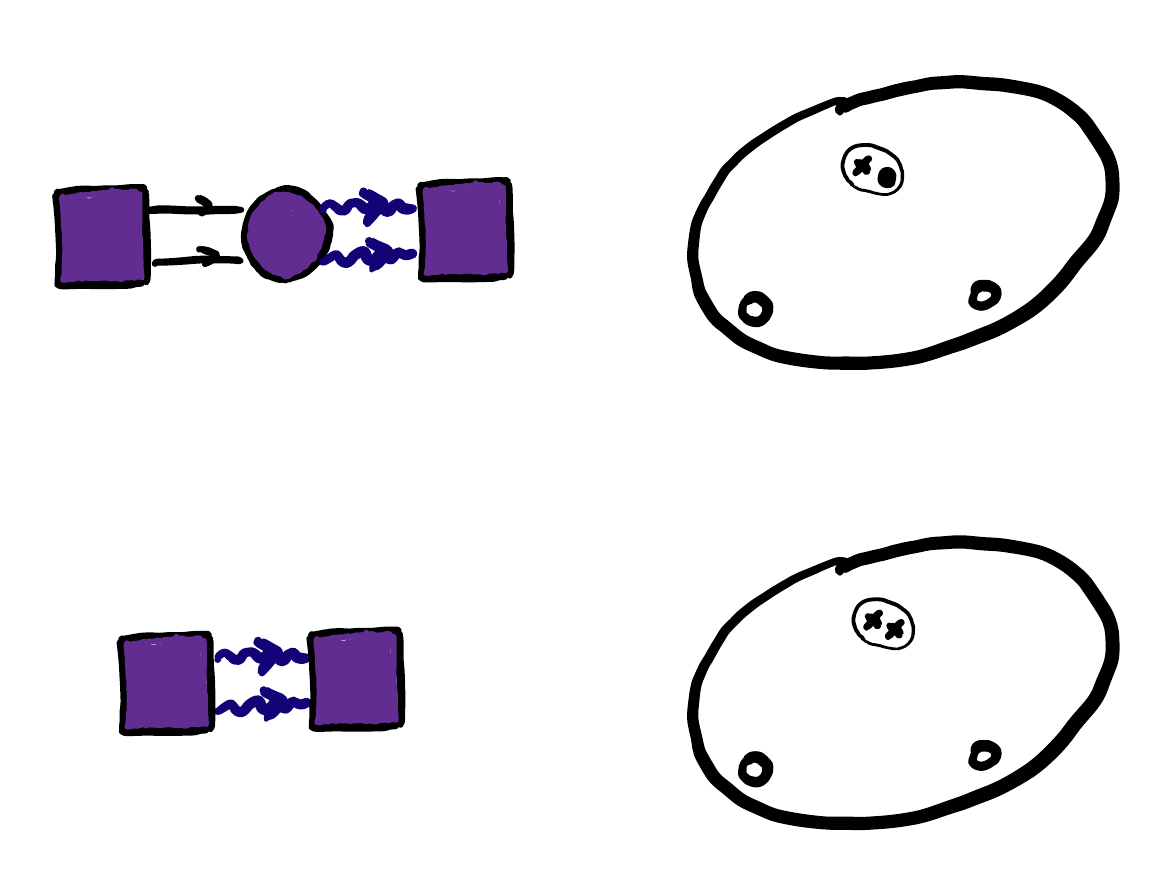}
\caption{Models one obtains by flows from a sphere with three punctures with $SU(4)$ symmetry. The black circle is  $\hat { SU  }   (2)   $ puncture. The white circle is $SU(4)$ puncture and crosses are basic punctures.}
\end{figure}
We can also compute the index and find that it is,
\be
&&1+q \, p\biggl( 3g-3+s+2s'+2s''+\sum_{j=1}^s (-{\bf 15}_j+{\bf 20}'_j)+\nonumber   \\
&&+\sum_{j=1}^{s'}(-{\bf 3}_j+{\bf 5}_j)\biggr)+\cdots  \;.\,
\ee On a general point of the conformal manifold all symmetry is broken and the dimension is then $3g-3+6s+4s'+2s''$.  We note that the empty puncture contributes to the dimension of the manifold and to the anomalies as one third of the $SU(4)$ puncture, and also $SU(2)$ punctures contribute as two thirds of the $SU(4)$ punctures.  Note that the number of exactly marginal couplings preserving the symmetries with the non $SU(4)$ punctures does not match the complex structure moduli. Similar phenomena is also known to occur for minimal punctures in some other examples of compactifications such as ${\cal N}=1$ class ${\cal S}$ theories, see \cite{Benini:2009mz}, so this is not necessarily alarming. We note however the following fact.

Let us assume that the theory has basic punctures such that the conformal manifold is given by complex structure moduli. We do not have direct evidence for such a statement, but it has some appeal, and we will see that assuming this some properties work out in a nice manner. 
Let us assume that the basic puncture with no symmetry is such that the empty puncture is a collision of two such punctures, and the $SU(4)$ puncture is a collision of six basic punctures. Moreover let us also  conjecture that the $SU(2)$ puncture we have obtained in the flow is a collision of the basic puncture and a new type of $SU(2)$ puncture, denoted $\hat {SU  }  (2)$. Denoting the number of the $SU(4)$ punctures $s$,  $\hat {SU   }    (2) $ punctures $s_p$, and  basic punctures $s_t$ the index is,
\be
&&1+q \, p\biggl( 3g-3+s+s_p+s_t+\sum_{j=1}^s (-{\bf 15}_j+{\bf 20}'_j)+\nonumber   \\
&&+\sum_{j=1}^{s_p}(-{\bf 3}_j+{\bf 5}_j)\biggr)+\cdots  \;.\,
\ee 
This is consistent with the geometric interpretation of all the conformal deformation   as associated to complex structure moduli. On generic points of the manifold all symmetries are broken and the punctures split to basic. We need to be careful with the twist of the punctures. If we assume that we associate a given twist to the $SU(4)$ puncture the flow does not change it. It then means that the basic puncture and the  $\hat{SU }
(2)$ puncture obtained in the flow have inverse twist. On the conformal manifold  $\hat{SU   }   (2)$ puncture of given twists can split into three basic punctures such that  two are of
same twist and one of inverse twist. Likewise, an $SU(4)$ puncture can split into two $\hat  { SU} (2)  $ punctures of inverse twist.

\noindent Note that starting with a sphere with three $SU(4)$ punctures, and closing one to $SU(2)$, we obtain a sphere with sixteen basic punctures. The dimension of the space of complex structure moduli is thirteen. Let us count the baryonic marginal operators of this theory which is just an $SU(4)$ gauge theory with eight flavors. We have seventy baryons and antibaryons and the gauge coupling as marginal operators. The symmetry is $SU(8)\times SU(8)$ and two $U(1)$ symmetries one of which is anomalous. The symmetry is completely broken on the conformal manifold, thus the number of exactly marginal operators is $70+70+1-63-63-1-1=13$. This agrees with the complex structure moduli. In this case we have also marginal operators coming from the mesons but they do not have a general six dimensional interpretation.

\

\noindent{\bf Acknowledgments}:~
We would like to thank  Michele Del Zotto, Hee-Cheol Kim, Guglielmo Lockhart, Yuji Tachikawa, and Futoshi Yagi for interesting conversations. We also would  like to thank Oren Bergman and Zohar Komargodski for collaboration at the early stages of the project and for many  important comments.
 The research of SSR was  supported by Israel Science Foundation under grant no. 1696/15 and by I-CORE  Program of the Planning and Budgeting Committee.
  GZ is supported in part by  World Premier International Research Center Initiative (WPI), MEXT, Japan.

\end{document}